\documentclass[twocolumn,showpacs,superscriptaddress]{revtex4}

\usepackage{dcolumn}
\usepackage{amsmath}

\makeatletter

\providecommand{\LyX}{L\kern-.1667em\lower.25em\hbox{Y}\kern-.125emX\@}

\usepackage[T1]{fontenc}
\usepackage[latin1]{inputenc}
\usepackage{graphics}
\usepackage{comment}

\makeatother

\begin{document}

\title{Coulomb breakup in a Transformed Harmonic Oscillator Basis}

\author{A.M.\ Moro}
\affiliation{Departamento de F\'{\i}sica At\'omica, Molecular y
Nuclear, Facultad de F\'{\i}sica, Universidad de Sevilla,
Apartado~1065, 41080 Sevilla, Spain}

\author{F.\ P\'erez-Bernal}
\affiliation{Departamento de F\'{\i}sica Aplicada, Universidad de Huelva, 21071
  Huelva, Spain}

\author{J. M. Arias}
\affiliation{Departamento de F\'{\i}sica At\'omica, Molecular y
Nuclear, Facultad de F\'{\i}sica, Universidad de Sevilla,
Apartado~1065, 41080 Sevilla, Spain}

\author{J.\ G\'omez-Camacho} 
\affiliation{Departamento de F\'{\i}sica At\'omica, Molecular y
Nuclear, Facultad de F\'{\i}sica, Universidad de Sevilla,
Apartado~1065, 41080 Sevilla, Spain}


\date{\today}

\begin{abstract}
The problem of Coulomb breakup in the scattering of a two-body loosely
bound projectile by a heavy target is addressed. A basis of
transformed harmonic oscillator (THO) wave functions is used to
discretize the projectile continuum and to diagonalize the Hamiltonian
of the two-body system. 
Results for the reaction
$^{8}$B+$^{58}$Ni  at subcoulomb energies are presented. 
Comparison of different
observables with those obtained with the standard Continuum
Discretized Coupled-Channels (CDCC) method shows good 
agreement between both approaches. 

\end{abstract}


\maketitle

\vspace{2cm}
The collision of a weakly bound system with a target  represents a 
challenging and interesting problem in quantum physics, including
atomic, molecular and nuclear physics. A proper understanding
of the process requires an appropriate treatment of the unbound part of the spectrum
of the loosely bound system. In nuclear collisions,  the problem was first 
addressed in the context of deuteron scattering. One of the first successful approaches   
to this problem was the Continuum Discretized Coupled-Channels (CDCC)
method, firstly introduced by Rawitscher \cite{Raw72}, and later developed and employed 
by  other groups \cite{Yah86,Kam86,Aus87}. Within the CDCC method, the reaction
process of a loosely two-body projectile and a heavy structureless target is treated 
within a three-body picture. The idea of the method is to represent the 
continuum part of the two-body projectile spectrum by a finite set of square integrable 
states. To this end, the continuum is divided 
into a finite set of energy intervals. For each interval, or bin, 
a representative function is constructed by superposition of true scattering wave functions
within that interval. By construction, the set of functions obtained in this way are 
normalizable and mutually orthogonal. By projecting the  
Schr\"odinger equation onto the bound and bin wave functions, a set of coupled equations is
obtained.  The method has been extremely successful and is a reliable reference
for any other alternative method. In spite of this, it has been criticized by some 
authors due to the apparent arbitrariness in the definition of the continuum bins 
\cite{Saw86}. Furthermore, the generalization of the method to treat the three-body continuum
is not obvious.

As an alternative to the binning procedure, other methods have been used to represent the 
continuum spectrum of a two-body system by a discrete and finite set of square 
integrable, or $L^2$, states such 
as Laguerre \cite{Rod97,Bray98} or Gaussian \cite{Mac87a,Hiy03} functions. A family of 
these states, usually called {\em pseudostates}, is used to diagonalize the 
Hamiltonian of the two-body system. The resulting 
eigenstates  are then used within the CDCC calculation in exactly the same way as the continuum
bins. 
In this context,
a new basis suitable for continuum discretization has been proposed and applied to 
the scattering of two-body composite systems in a series of recent 
works \cite{Per01b,Per01,Mor02,MPRAG02,PMAG03,Ro04}. The method generates
a discrete representation of the continuum spectrum starting from 
the ground state wave function, which is the only needed input. A
local scale point transformation (LST)
\cite{SP88,PS91} that transforms this function into the
harmonic oscillator ground state wave function is defined. Once the LST is
obtained, the inverse transformation produces from the harmonic
oscillator states the wave functions that represent the continuum (and
other bound states if they exist) of the two-body system. The
corresponding pseudostate set is known as the Transformed Harmonic Oscillator (THO) basis.
The method was first developed in Ref.~\cite{Per01b}  
for simple one dimensional problems 
and later extended in order to check its applicability and limitations
\cite{Per01,Mor02,MPRAG02,PMAG03,Ro04}.
In particular, in \cite{Mor02} it was shown that the combination of the THO 
discretization method 
with the coupled channels technique, named CDCC-THO, can be useful to describe continuum
effects in nuclear collisions. In \cite{Mor02}, several restrictions
were imposed for the sake of
simplicity: the deuteron was taken to be a 
pure \(s\)-state, the effect of Coulomb breakup was neglected, and 
the nuclear interaction of the proton and the neutron with the target
was accepted to induce the coupling just to \(s\)-wave
breakup states. Moreover, the deuteron ground state was 
represented by an analytical wave function.
Within this simplified scenario, it was 
shown that the method was in excellent agreement with
the standard CDCC method. 

In this work we apply the CDCC-THO method 
to the reaction $^{8}$B+$^{58}$Ni at 25.8 MeV, in which the  $^{8}$B projectile
is modelled as a weakly bound two-body system, proton+$^{7}$Be. This reaction has been 
extensively analyzed within the standard CDCC formalism \cite{Nun99,Tos01} and 
the CDCC with a Gaussian basis \cite{Ega04}. The motivations of this 
work are the following. First we would like to show that the method
can be equally implemented in the usual situation in which the
ground state of the system is not known analytically but is calculated
numerically. Second, we aim to show that the method can be generalized in a 
straightforward way to describe the continuum part of the spectrum for
arbitrary partial waves. 
Since the $^{8}$B ground state corresponds mainly to a $p-$wave configuration, the 
THO method provides a discrete representation for the $\ell=1$
continuum. We will show below, however, that it can be easily extended to generate also a 
representation for the $\ell=0$  
continuum. Finally, we will show
that, despite the finite extension and exponential asymptotic behaviour of the 
THO basis, these states are suitable to describe situations dominated 
by long range interactions, such as the dipole Coulomb couplings arising
in collisions of a loosely bound projectile by  heavy ions. 

For that purpose,
we first present a brief review of the the basic formulation of the THO method and 
describe how it can be 
extended to different continuum configurations. Then, we present
calculations, in which the THO wave functions are used for a coupled
channels calculation (CDCC-THO),
to show the convergence of the method for several magnitudes related
to the $^{8}$B+$^{58}$Ni reaction at 25.8 MeV. We use the standard
CDCC calculations as a reference to compare with.


The standpoint of the THO 
method is the ground state wave function of the two-body system, here denoted as 
$\phi_{b,\ell_0}(r)$. The subscript $\ell_0$ represents the inter-cluster relative angular momentum.
To simplify the notation, we do not consider the intrinsic spins of
the fragments. 
By use of a local scale 
transformation (LST) \cite{SP88,PS91}, this wave function is converted into a 
harmonic oscillator (HO) wave function. Thus, the function $s(r)$ defining
the LST is given by
\begin{equation}
\label{eq:LST}
\phi _{b,\ell_0}(r)=\sqrt{\frac{ds}{dr}}\phi ^{HO}_{0,\ell_0}(s(r)).
\end{equation}
where $\phi ^{HO}_{0,\ell_0}(s)$ is the radial part of the HO wave function 
for the orbital angular momentum $\ell_0$.
Once the $s(r)$ function has been obtained, the THO basis is generated  
by applying the same LST calculated for the
ground state to the rest of HO wave functions.
Due to the simple analytical structure of the harmonic oscillator
wave functions, this is equivalent to multiply the ground state function
by the appropriate Laguerre polynomials $L_{n}^{\ell +1/2}(s)$ \cite{Mor02}
\begin{equation}
\label{eq:tho}
\phi ^{THO}_{n, \ell}(r)=\left[ s(r)\right]^{\ell-\ell_0} 
L_{n}^{\ell + 1/2}(s(r)^2)\phi _{b, \ell_0}(r).
\end{equation}

Notice that, by construction, the family of functions  
\( \phi ^{THO}_{n, \ell}(r) \) are orthogonal
and constitute a complete set. Moreover, they decay exponentially
at large distances, thus ensuring the correct asymptotic behaviour
for the bound wave functions. However, in general, these functions  are not
eigenstates of the internal Hamiltonian. Then, the following step is to diagonalize
the Hamiltonian using a truncated THO basis. As a result of the diagonalization
a new set of functions, denoted \( \{\phi ^{N}_{j, \ell}(r);\, j=0,\ldots ,N\} \),
with eigenvalues \( \epsilon _{0},\ldots ,\epsilon _{N} \) are obtained.
Here, \( N+1 \) is the number of functions retained in the THO basis,
\( j=0 \) standing for the ground state. Thus 
\( \phi ^{N}_{0,\ell_0}(r)=\phi ^{THO}_{0,\ell_0}(r)=\phi _{b,\ell_0}(r) \)
and so \( \epsilon _{0}=\epsilon _{b} \), while the rest of eigenstates
constitute our representation of
the bound and unbound energy spectrum. Those eigenstates  
at negative energy will represent of bound states of the system. In particular, 
for $\ell < \ell_0$, the diagonalization generates
negative energy states, which correspond to occupied states. These states are Pauli forbidden
states, that should be excluded from the calculations. 

In \cite{Mor02}, the transformation given by Eq.~(\ref{eq:tho}) was defined only for 
continuum states with the same angular momentum $\ell$ as the ground state. In 
the present work, we generalize this procedure by applying the same transformation
to obtain THO states for $\ell \ne \ell_0$. In particular, we will apply the 
method to describe the $\ell=0$ and $\ell=1$  continuum states of the $^7$Be+p  
system.

To simplify the calculations, we neglect the $^7$Be and valence proton 
spins. Therefore, the  $^8$B ground state is given by a $\ell=1$, $J=1$ 
configuration. Using Eq.~(\ref{eq:tho}) we generate the THO basis for 
the $\ell=0$ and $\ell=1$ states. Then, we retain a finite number of
states and diagonalize the $^8$B Hamiltonian in this truncated space. The
resulting eigenvalues and their associated wave functions constitute a discrete 
and finite representation of the continuum for the $^7$Be+p system.

The bound and continuum states are used to generate the diagonal and 
non-diagonal coupling potentials that enter the system of coupled equations. 
In both the standard CDCC and CDCC-THO methods, these coupling 
potentials are defined as 
\begin{equation}
U_{\alpha,\alpha'}(R)=
\langle \phi_{n, \ell} | U_\mathrm{[p-Ni]} + U_\mathrm{[7Be-Ni]} |
 \phi_{n', \ell'} \rangle ,
\end{equation}
with $\alpha = \{n,\ell\}$,  $\alpha' = \{n',\ell'\}$. 
The internal wave functions $\phi_{n, \ell}(r)$ are 
represented by either continuum bins
or the THO basis, depending on the discretization method. 
The proton-$^7$Be nuclear interaction was 
represented by a simple Woods-Saxon form with the potential 
parameters of Esbensen and  Bertsch \cite{Esb99}. This interaction 
is used to generate the ground state wave function for the $^8$B system 
as well as for the diagonalization of the  $^8$B Hamiltonian in the THO basis. The 
proton-$^{58}$Ni interaction was taken from the parametrization 
of Becchetti and Greenless \cite{Bec69}. For the  $^7$Be-$^{58}$Ni 
system, we used the interaction of Moroz {\em et al.} \cite{Moroz82} 
following the choice
of other authors \cite{Ega04,Tos01}. 
All possible couplings ($s-s$, $s-p$ and 
$p-p$) were included in the calculations for multipolarities 
$\lambda$=0,1 and 2. 
In the standard CDCC calculation, 
the continuum spectrum was discretized into energy bins of equal 
momentum width and up to a maximum excitation energy of 
$\epsilon_{\max}=10$ MeV. In particular, we used $N=16$ bins for $\ell=1$ and
 $N=32$ for  $\ell=0$. 
The bin wave functions were calculated up
to 100 fm. For a meaningful comparison, we
take the same potential parameters as in the CDCC-THO calculation. 
The coupled equations were solved for projectile-target orbital angular momenta
up to $L_{\max}=500$ 
and integrated up to a radius  $R_{\max}=500$ fm. 

\begin{figure}[t]
{\par\centering \resizebox*{0.25\textwidth}{!}
{\includegraphics{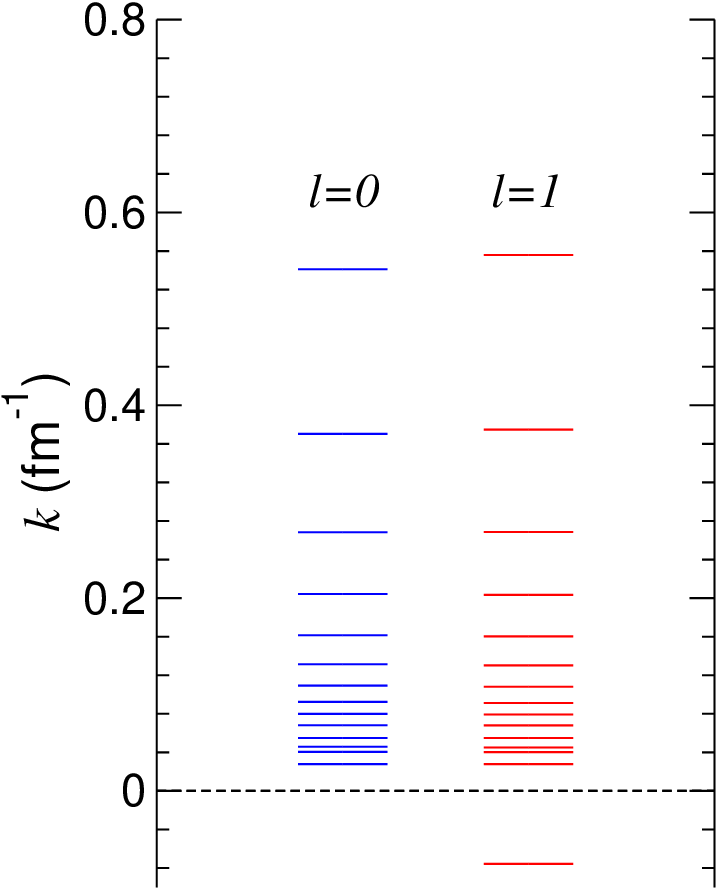}}\par}
\caption{\label{fig:ener} Eigenvalues obtained upon diagonalization
of the internal Hamiltonian in a THO basis with $N=20$ states for 
$s$ and $p$-waves. The vertical axis corresponds to the proton-$^7$Be 
linear momentum. 
The $p$-state at negative energy corresponds
to the ground state.}
\end{figure}

In Fig.~\ref{fig:ener} we represent the eigenvalues obtained upon 
diagonalization of the $^8$B Hamiltonian in a THO basis with $N=20$ states.
The vertical axis corresponds to the linear momentum  $k_i=\sqrt{2 \mu  \epsilon_i}/\hbar $, 
where $\mu$ is the proton-$^7$Be reduced mass.
It should be noticed that these eigenvalues tend to concentrate close to zero energy. As
the excitation energy increases, the energy distribution becomes more sparse.  It is 
interesting to 
note that a deeply bound state ($\epsilon \approx -15$ MeV) appears for 
$\ell=0$ (not shown in Fig.~\ref{fig:ener}). This state corresponds
to the $1s$ level of the $^8$B nucleus. Since this state is
occupied it is  removed from our model space. We note also that some eigenvalues may lie at 
very high energies. In principle, one could include these states in the system of the coupled 
equations. In practice, however, states at very high energies are very weakly coupled to 
the ground state and, hence, do not influence the dynamics of the reaction. Besides, the inclusion
of these states makes the calculation computationally more intensive and 
may even lead to convergence problems. For these
reasons, in our calculations those states above a certain excitation energy are completely removed from
the coupled equations. This maximum energy will depend, indeed, on the particular reaction. In 
this case, we found that states above 10 MeV have no effect whatsoever on the
reaction observables we are studying and, therefore, only those states
with excitation energies below this value are retained in our
calculations. Notice that states above 10 MeV have been also omitted from Fig.~\ref{fig:ener}.

\begin{figure}[t]
{\par\centering \resizebox*{0.45\textwidth}{!}
{\includegraphics{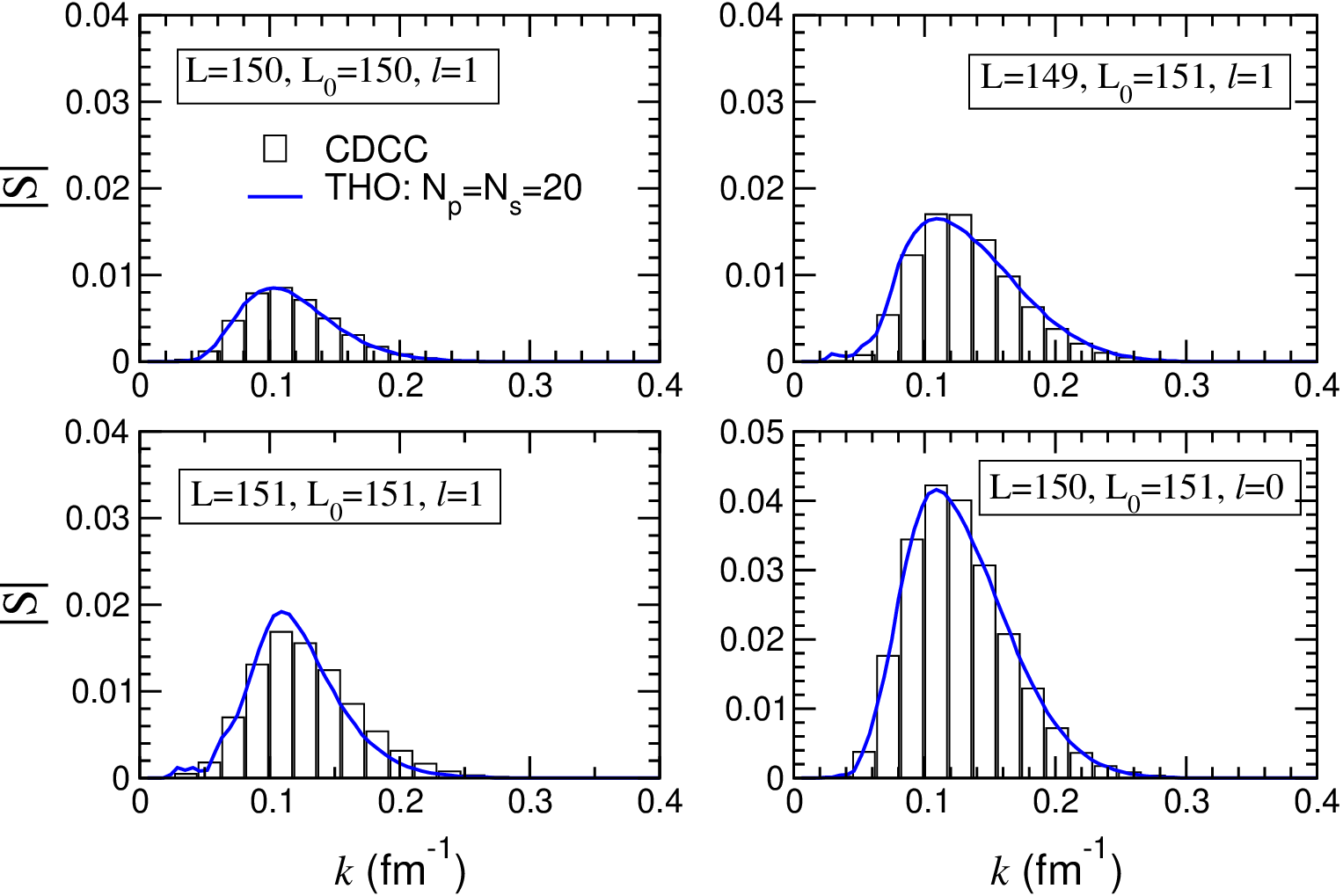}}\par}
\caption{\label{sbu_sp_j150} Breakup S-matrix elements for the 
total angular momentum $J=150$
as a function of the asymptotic $^7$Be+p relative momentum.}
\end{figure}

In Fig.~\ref{sbu_sp_j150} we plot the modulus of some components of the
breakup $S-$matrix for a total angular momentum $J=150$ which, assuming a
classical trajectory, corresponds to a scattering angle of 10$^{\circ}$. As 
noted in \cite{Ega04} at these angles breakup is mostly due to the Coulomb
interaction. In the standard CDCC, the breakup $S-$matrix is obtained by dividing
the discrete $S-$matrix to the continuum bins by the square root of the bin width, $\Delta k_i$. In 
the CDCC-THO method, one could apply a similar procedure by assigning a width to each pseudostate. In 
\cite{Mor02} we used this approach  to calculate  the differential 
breakup cross section from the cross section to individual pseudostates. In particular, we assumed
that the width of the $i$th pseudostate is approximately given by 
$\Delta_i=(\epsilon_{i+1}-\epsilon_{i-1})/2$. In this work, we adopt a more sophisticated approach,
previously proposed in \cite{Mat03} in the context of the Gaussian expansion pseudostate method. In this method,
the breakup $S-$matrix elements $S_{\alpha':\alpha}(k)$, which depend 
on the continuous variable $k$, as well as on the initial and final angular momenta,
are obtained by an appropriate superposition of  the discrete $S-$matrix elements  
$\hat{S}_{\alpha':\alpha}(k_i)$ resulting from 
the solution of the coupled-channels equations, as \cite{Mat03}
\begin{equation}
\label{scont}
S_{\alpha':\alpha}(k) 
\approx
\sum_{j=1}^{N} 
\langle \phi^{(s)}_{k,\ell} | \phi_{j, \ell} \rangle 
\hat{S}_{\alpha':\alpha}(k_i)
\end{equation}
where  $\phi^{(s)}_{k,\ell}(r)$ is the true scattering wave function for the partial 
wave $\ell$ and energy 
$\epsilon=\hbar^2 k^2/2\mu$. The sum runs over the set pseudostates included in the 
coupled-channels calculation.
The indexes $\alpha$ and $\alpha'$ denote the initial  and final channels, that is,  
$\alpha =\{gs; L_0, \ell_0, J\}$ and $\alpha' =\{i; L, \ell, J\}$, where $L_0$ ($L$) is the 
initial (final) projectile-target orbital angular momentum.

The histogram represented  in Fig.~\ref{sbu_sp_j150}  corresponds to the 
standard CDCC calculations, and the lines to the CDCC-THO results calculated with 
a THO basis with $N_p$=$N_s$=20 states. We verified that increasing the number of basis
states does not change the calculated $S-$matrices, thus indicating the convergence of the THO
method with respect to the basis size.
Furthermore, the THO calculation with a  $N_p$=$N_s$=10 states gives already a
result very close to result presented in Fig.~\ref{sbu_sp_j150}.
The fast 
convergence of the THO in this case can be attributed to the fact that the
Coulomb interaction tends to populate mainly states in the continuum at low
excitation energies, where the density of THO states is higher.

\begin{figure}
{\par\centering \resizebox*{0.35\textwidth}{!}
{\includegraphics{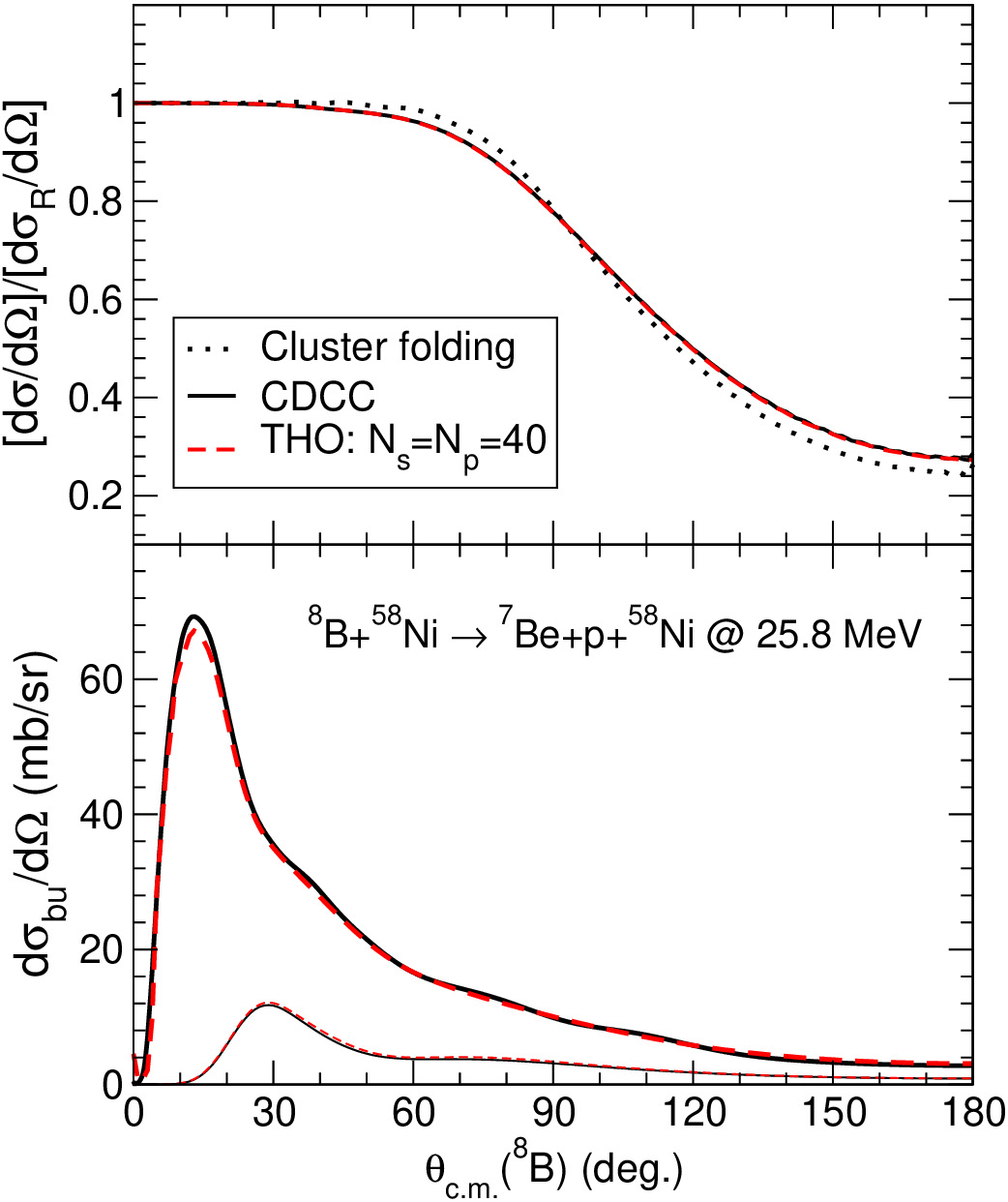}}\par}
\caption{\label{b8ang} Elastic (upper) and breakup (bottom) angular 
distribution  for the $^8$B+$^{58}$Ni reaction at 25.8 MeV in a model
space including $s-$ and $p-$wave states for the  $^7$Be+p continuum. In
both plots, the thick solid line represents the converged CDCC calculation
while the thick dashed line corresponds to the CDCC-THO calculation. 
The same potential parameters and partial waves are used for both calculations. 
The dotted
line in the upper panel is the pure cluster-folding calculation without continuum.
The thin lines in the bottom panel
are the calculations including only nuclear breakup. }
\end{figure}

In Fig.~\ref{b8ang} we represent the elastic (upper panel) and 
breakup (lower panel) angular distributions. The standard CDCC calculation
is represented in both panels by a thick solid line. The thick dashed lines represent
the CDCC-THO calculation
with $N=40$ states which, as can be seen, is almost identical to the standard CDCC  
calculation. The dotted line in the upper panel corresponds to the elastic 
calculation omitting the coupling to the continuum. Since both the CDCC and CDCC-THO use the same
ground state wave function and interactions, this calculation is identical in both methods.
The thin lines in the bottom panel are the calculated breakup angular
distributions including only nuclear breakup.

\begin{figure}
{\par\centering \resizebox*{0.35\textwidth}{!}
{\includegraphics{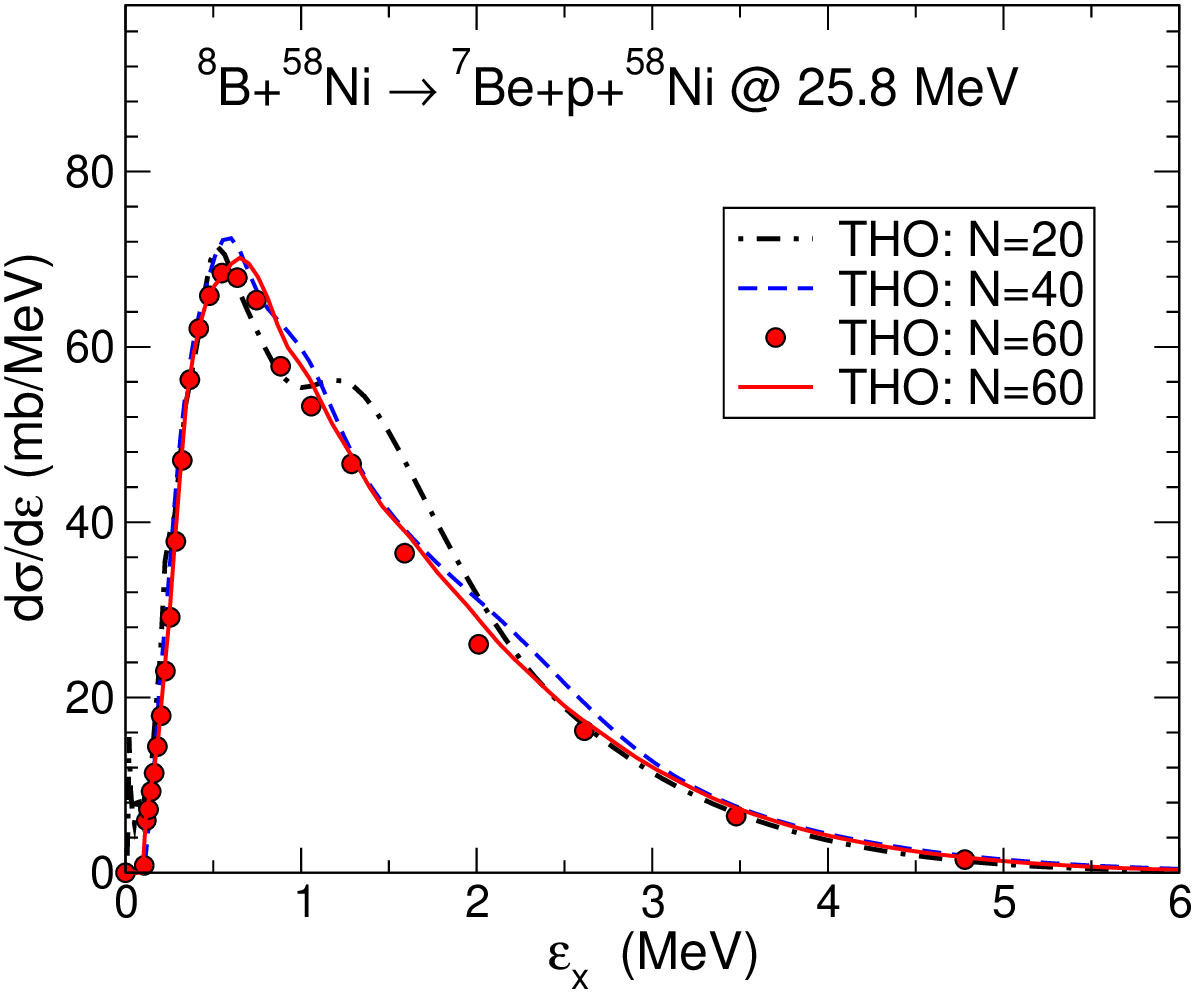}}\par}
\caption{\label{penerdist} Breakup  cross section to $p$ states as a function of the 
$^8$B internal excitation energy and
integrated from zero to 180 degrees.
The lines correspond to the CDCC-THO calculations with different basis dimensions, 
evaluated from continuous $S$-matrices according to  Eq.~(\ref{scont}). The solid circles represent
the THO calculation with $N_s=N_p=60$ states, in which the cross section to each pseudostate is 
divided by the estimated energy width (see text).}
\end{figure}

Next, we study the breakup differential energy cross section,  $d\sigma/dE$. 
In the standard
CDCC method, a natural way to approximate this quantity is to divide the cross section for each final state 
by the bin width \cite{Yah86,Tos01}. In the THO method, we could proceed in a similar way, by defining a width 
for each pseudostate. Alternatively, one can calculate the breakup cross section from the continuous breakup 
$S-$matrices evaluated by means of Eq.~(\ref{scont}). Both methods are compared below. 

In  Fig.~\ref{penerdist}, we depict the  differential energy cross section to $\ell=1$ continuum states obtained
within the CDCC-THO method with several basis sizes, as indicated by the labels. The  lines 
were obtained using the continuous $S$-matrices given by Eq.~(\ref{scont}) for different 
basis dimensions. The circles represent the 
calculation with $N_s=N_p=60$ states  in which  $d\sigma/dE$ is approximated by  
dividing the cross section of the $i$th pseudostate  by the
width $\Delta_i=(\epsilon_{i+1}-\epsilon_{i-1})/2$.  We see that both
procedures to obtain $d\sigma/dE$  give similar results. From this 
figure we can also infer that the calculation converges as the number of basis states is increased. 
We notice
that, at low excitation energies, the method converges even with a small basis. From this result we can conclude
that the method is very suitable to describe Coulomb breakup reactions. 
Convergence of the breakup cross section for excitation energies above 1 MeV requires a 
relatively large THO basis. 
This slower convergence rate is due to the fact that  the density of THO states diminishes as the excitation 
energy increases, and more states are required to describe the excitation into this region.
As the number of basis states increases, the solution of the coupled equations becomes slower. 
However, we note 
that the actual number of states included in the coupled equations is reduced with respect to the initial 
basis dimension. 
As explained above, after the
diagonalization only those states with excitation energy below 10 MeV are included in the coupled-channels 
calculation. In addition, the number of basis states  is further reduced by removing those states with excitation 
energy below $\sim 0.1$ MeV. Once these two restrictions are applied, 
the number of
continuum states actually included in the  
coupled-channels calculations for the case $N=60$ is $N_\mathrm{coup}=27$ for both the  $s$ and $p$ waves. We studied also 
the convergence of the method for the breakup 
to $s$-states. We found that, in this case, convergence is achieved with a smaller number of states,
compared to the $p$-wave breakup.

\begin{figure}
{\par\centering \resizebox*{0.35\textwidth}{!}
{\includegraphics{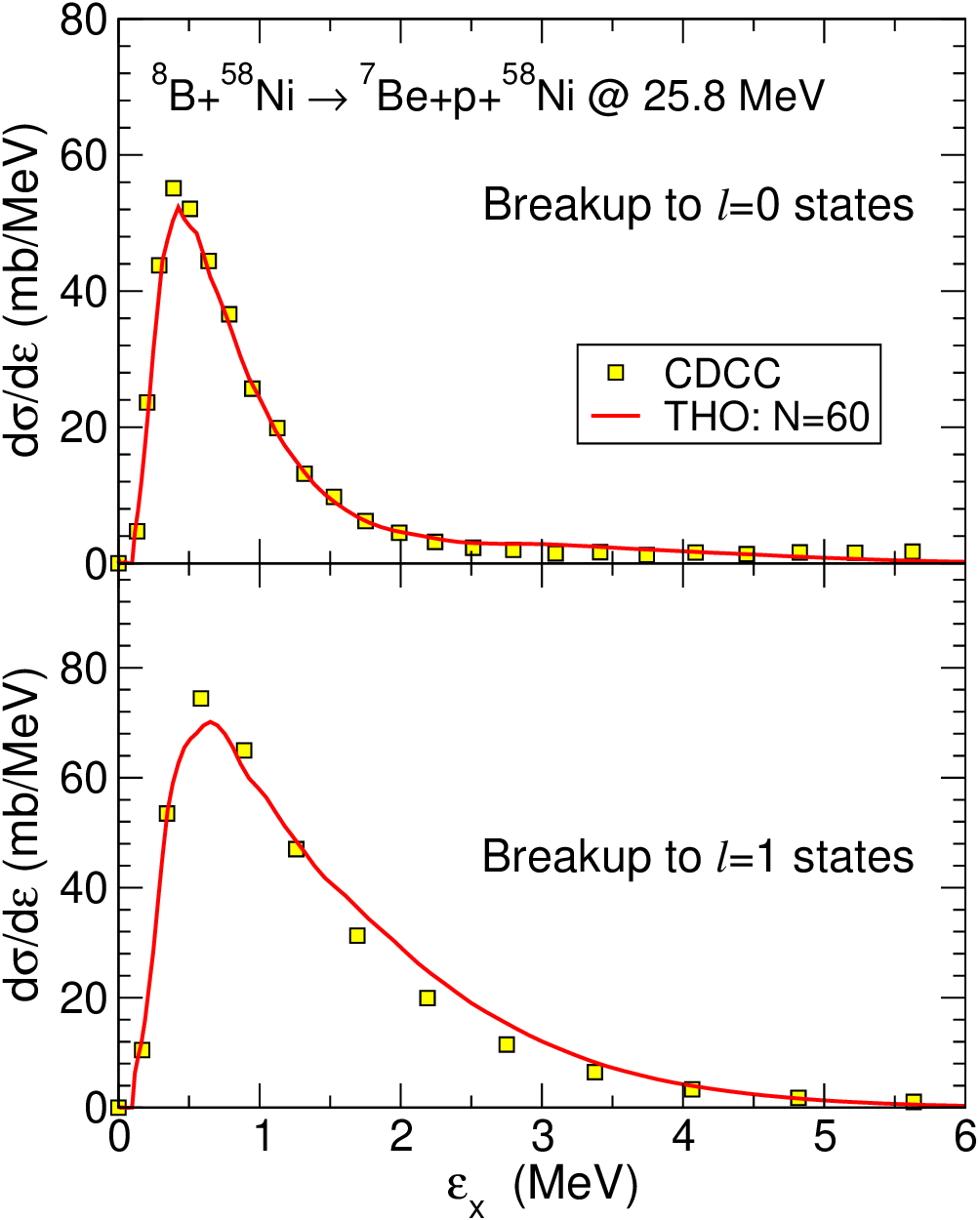}}\par}
\caption{\label{enerdist} Breakup  cross section as a function of the 
 $^8$B internal excitation energy for the  $^8$B+$^{58}$Ni reaction at 
25.8 MeV integrated from zero to 180 degrees.
The squares correspond to the standard CDCC calculations, 
whereas the solid lines
represent the CDCC-THO calculation with a basis with  $N_s=N_p=60$ states. The separated 
contribution of $s$ (upper panel) and $p$ (lower panel) waves is displayed.}
\end{figure}

Finally, in Fig.~\ref{enerdist} we compare the CDCC (squares) and CDCC-THO (solid lines) 
breakup energy distributions. 
The latter corresponds to a basis
with  $N_s=N_p=60$ states for which, as we have shown above, good convergence of the THO method is
achieved.  We present separately the 
contribution of the $\ell=0$  (upper panel) and $\ell=1$ (bottom panel) continuum states. 
We see that both energy distributions  
are in good  agreement  with the CDCC calculation. 

In conclusion, in this work we have presented an extension of the THO method, 
formerly applied to describe the $\ell=0$ deuteron continuum, to generate 
a discrete representation of the continuum of  a two-body system for any partial wave. As
in the original formulation, the only {\it a priori} prerequisite for the application
of the method is the knowledge of the ground state wave function of the system, either
analytically or numerically. A local scale transformation, $s(r)$, is 
then defined such that
it converts the ground state wave function of the system into the harmonic 
oscillator ground state. Then, the THO basis is obtained by multiplying the ground
state wave function by a set of Laguerre polynomials expressed in the variable $s(r)$. 
This THO basis constitutes a discrete and finite representation of the continuum part of
the  spectrum with the same inter-cluster angular momentum as the ground state. For other
partial waves, the method is generalized in a straightforward way by simply applying 
the same scaling transformation. The method has been applied to the reaction 
$^8$B+$^{58}$Ni at subcoulomb energies, showing a fast convergence with the respect to the
number of  basis states. Our results agree very well with those obtained with  
the standard CDCC method. In a recent work, we have successfully
extended the method to describe the  continuum of a three-body system 
\cite{manoli05} and the 
application of the method to describe the collision of a three-body loosely bound
system by a target is underway.

\acknowledgments
This work has been partially supported by the Spanish Ministerio de
Educaci\'on y Ciencia and by the European regional development fund
(FEDER) under projects number FIS2005-01105, FPA2005-04460 and FPA2003-05958.
A.M.M. acknowledges a research fellowship from the Junta de 
Andaluc\'{\i}a.


\bibliographystyle{apsrev}
\bibliography{thob8}

\end{document}